\documentclass[twocolumn,preprintnumbers,amsmath,amssymb]{revtex4}

\usepackage{graphicx}
\usepackage{dcolumn}
\usepackage{bm}

\bibstyle{apsrev}

\begin{document}


\title{The radiative lifetime of metastable CO ($a^3\Pi, v=0$)}

\author{Joop J. Gilijamse}
\author{Steven Hoekstra}
\author{Samuel A. Meek}
\author{Markus Mets\"al\"a}
\author{Sebastiaan Y.T. van de Meerakker}
\author{Gerard Meijer}
\email{meijer@fhi-berlin.mpg.de} \affiliation{Fritz-Haber-Institut
der Max-Planck-Gesellschaft, Faradayweg 4-6, 14195 Berlin, Germany}
\author{Gerrit C. Groenenboom}
\email{gerritg@theochem.ru.nl}
\affiliation{Theoretical Chemistry, Institute for Molecules and Materials, Radboud University Nijmegen,
Toernooiveld 1, 6525 ED Nijmegen, The Netherlands}
\date{\today}

\begin{abstract}
We present a combined experimental and theoretical study on the
radiative lifetime of CO in the $a^3\Pi_{1,2}, v=0$ state. CO
molecules in a beam are prepared in selected rotational levels of
this metastable state, Stark-decelerated and electrostatically
trapped. From the phosphorescence decay in the trap, the radiative
lifetime is measured to be $2.63\pm0.03$~ms for the $a^3\Pi_1, v=0,
J=1$ level. From spin-orbit coupling between the $a^3\Pi$ and the
$A^1\Pi$ state a $20\%$ longer radiative lifetime of 3.16~ms is
calculated for this level. It is concluded that coupling to other
$^1\Pi$ states contributes to the observed phosphorescence rate of
metastable CO.
\end{abstract}

\maketitle

Triggered by the observation of CO in the upper atmosphere of Mars
with a UV spectrometer on board the Mariner 6 spacecraft in 1969
\cite{Barth:1969}, researchers started studying the UV bands of CO
in increasing detail. In particular, the $a^3\Pi\leftarrow
X^1\Sigma^+$ transition between the ground state and the lowest
electronically excited state of CO, the so-called Cameron bands,
became the subject of theoretical as well as experimental interest.
From the appearance of the spectrum it was concluded that the
intensity in this spin-forbidden transition mainly originates from
spin-orbit mixing of the $a^3\Pi$ state with a $^1\Pi$ state; a detailed analysis revealed that mixing of
$^1\Sigma$ states with the $a^3\Pi$ state contributes less than 1\%
to the total intensity \cite{James:JMS:1971}. Using perturbation
theory and taking only the interaction with the $A^1\Pi$ state into
account, James calculated the radiative lifetimes for various
ro-vibrational levels in the $a^3\Pi$ state \cite{James:1971}. These
lifetimes are strongly $J$-dependent, but the ratio of the lifetimes
of different $J$-levels is known with spectroscopic accuracy. The
$J=1$ level in the $a^3\Pi_1, v=0$ manifold, from now on indicated
as the ($\Omega$,$v$,$J$)=(1,0,1) level, has the shortest lifetime,
calculated by James as 2.93~ms; the (2,0,2) level, for instance,
lives $54.66$ times longer.

An accurate experimental value for the radiative lifetime of
ro-vibrational levels of metastable CO is a benchmark for
theoretical calculations on spin-orbit mixing of electronically
excited states as well as on transition dipole moments. It has proved to be
difficult, however, to experimentally determine the lifetime of any
of the levels of the $a^3\Pi$ state of CO accurately \cite{Johnson,
Lawrence, Slanger, Borst}. These lifetimes can either be extracted
from absorption measurements on the Cameron bands or, more directly,
from measurements of the phosphorescence decay. For absorption
measurements, experimental parameters like the line-integrated
number density of ground-state CO molecules in a certain
ro-vibrational level, together with the absorption line-shape and
the spectral profile of the light source have to be accurately
known. More problematic is that the Franck-Condon factors for the
vibrational bands of this transition also have to be known to be
able to deduce a radiative lifetime. For phosphorescence decay
measurements, on the other hand, the ro-vibrational state
distribution of the metastable molecules needs to be known, and, on
the timescale of the phosphorescence, collisions need to be avoided.
To achieve this, laser excitation of CO to single ro-vibrational
levels of the metastable state in the collision-free environment of
a molecular beam has been used
\cite{Jongma:JCP107:7034,Sykora:1999,Sykora:2000}. An intrinsic
problem in these molecular beam experiments is, however, that the
molecules move with a high speed, typically limiting the time during
which the phosphorescence can be detected to a fraction of a
millisecond.

Stark deceleration of a beam of polar molecules followed by
electrostatic trapping enables the observation of state-selected
molecules for several seconds \cite{Hoekstra:PRL98:133001}.
Recently, the radiative lifetime of vibrationally excited OH
radicals was determined by recording the temporal decay of their
population in the trap \cite{Meerakker:PRL95:013003}. Here we report
the Stark deceleration and electrostatic trapping of metastable CO
molecules that are laser prepared in either the (1,0,1) or in the
(2,0,2) level. The radiative lifetimes are measured by monitoring
the phosphorescence decay of the trapped molecules. Calculations to
rationalize the observed lifetimes are presented.

\begin{figure}[!ht]
\includegraphics[width=\columnwidth]{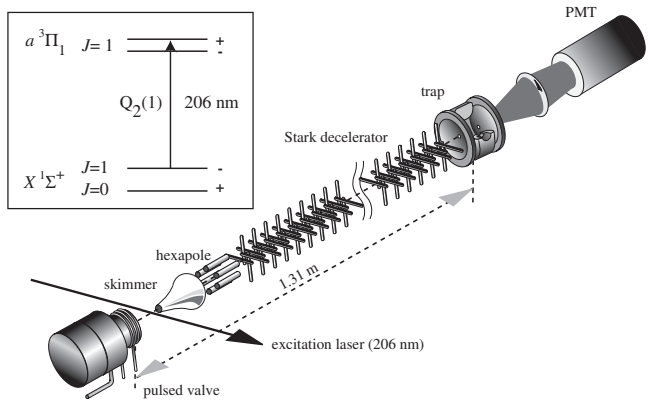}
\caption{\label{fig:set} Scheme of the experimental setup. A pulsed beam
of CO molecules, laser prepared in either the (1,0,1) level
(excitation scheme given in the inset) or the (2,0,2) level,
is slowed down in the Stark decelerator and stored in an electrostatic
trap. The phosphorescence to the electronic ground state is measured by
a photomultiplier tube (PMT).}
\end{figure}

Stark deceleration of metastable CO molecules has been described in
detail before \cite{Bethlem:IRPC22:73}. In the present experiment,
we use a Stark decelerator with 108 electric field stages to load
the metastable CO molecules in an electrostatic trap. The
experimental setup is schematically shown in Fig.\ \ref{fig:set}. A
pulsed beam of CO molecules with a mean velocity of 320 m/s is
produced by expanding a mixture of $20\%$ CO in Xe from a cooled
pulsed valve ($T=203$~K). A packet of CO molecules in the upper
(low-field seeking) $\Lambda$-doublet component of the (1,0,1) or
(2,0,2) level in the metastable $a^3\Pi$ state is prepared by
direct laser excitation from the $X^1\Sigma^+, v=0$ ground state.
After passing through a skimmer, the packet of molecules is slowed
down in the Stark decelerator and subsequently loaded and confined
in an electrostatic quadrupole trap. When in the (1,0,1) level,
about 10$^5$ CO molecules are trapped at a density of 10$^8$/cm$^3$
and at a temperature of around $20$~mK; in the (2,0,2) level an order
of magnitude less molecules are trapped at a somewhat higher
temperature. A detailed description of the Stark decelerator and of
the electrostatic trap that have been used can be found elsewhere
\cite{Meerakker:ARPC57:159}.

\begin{figure}[!ht]
\includegraphics[width=0.75\columnwidth]{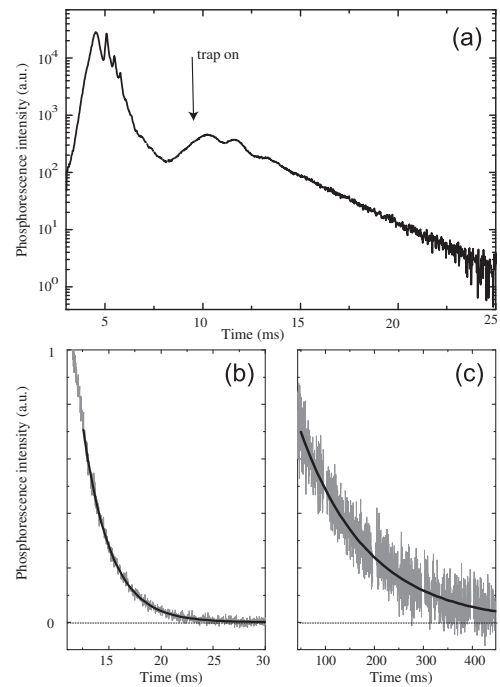}
\caption{\label{fig:dec} Semilogarithmic plot of the phosphorescence intensity
from the trap region as a function of time after production of CO molecules in
the (1,0,1) level in the source chamber (a). The time at which the trap is switched on is
indicated by an arrow. Phosphorescence decay curves are shown on a linear scale
for CO molecules in the (1,0,1) level (b) and in the (2,0,2) level (c). The solid lines are the results of a weighted
least-square fit to a single exponentially decaying curve.}
\end{figure}

The ultraviolet phosphorescence back to the electronic ground state
that escapes through a hole in one of the trap electrodes is imaged
with a lens onto a photomultiplier tube (PMT). In the upper part of
Fig.\ \ref{fig:dec} a semilogarithmic plot of the phosphorescence
signal is shown as a function of time. The CO molecules are prepared
in the (1,0,1) state at $t=0$~ms, and the strong phosphorescence
peak around 4-5~ms results from molecules that pass through the trap
with (more or less) the initial beam velocity. At 9.3~ms, the
decelerated packet of molecules arrives in the trap center and the
trap is switched on. After some initial oscillations, caused by the
collective motion of the molecules in the trap, an exponentially
decaying phosphorescence signal is observed. This signal is shown on
a linear scale in panel~$(b)$ of Fig. \ref{fig:dec}. A weighted
least-squares single exponential fit to this data in the
time-interval from 13 to 38~ms gives a decay time of
$2.63\pm0.03$~ms, where the error is an estimate accounting for the
systematic effect of the initial oscillations. The phosphorescence
of the trapped molecules is measured in the presence of the trapping
fields, i.e. in an electric field ranging from zero to 10~kV/cm.
Under these conditions the opposite parity states of the
$\Lambda$-doublet are mixed.

For the measurement of the phosphorescence decay of CO molecules in
the (2,0,2) level, shown in panel (c) of Fig.\ \ref{fig:dec}, the
repetition frequency of the molecular beam experiment is reduced
from the normally used 10~Hz to 2~Hz to allow for a $\sim500$~ms
observation time. To avoid detection of stray-light from the
excitation laser, that keeps on running at 10~Hz, no phosphorescence
signal is recorded during a 6~ms interval around the times that this
laser is fired. A single exponential fit to the data in the 50 to
450~ms interval yields a decay time of $140\pm4$~ms.

Apart from the phosphorescence to the ground state, optical pumping
by blackbody radiation and collisions with background gas can lead
to a decay of the signal of trapped molecules; at the present
densities, collisions between trapped molecules do not play a role.
Optical pumping due to room temperature blackbody radiation is
calculated to proceed at a rate of around 0.014~s$^{-1}$ for CO in
the $a^3\Pi, v=0$ state \cite{Hoekstra:PRL98:133001}. The
contribution of optical pumping to the overall trap loss rate can
therefore safely be neglected. The loss rate due to background
collisions was measured to be 0.17~s$^{-1}$ for trapped ground-state
OH and OD radicals in this apparatus. As these measurements were
done under otherwise identical conditions, we assume the same
collisional trap loss rate for metastable CO. For CO molecules in
the (1,0,1) level the loss due to collisions can then also be
neglected; for molecules in the (2,0,2) level collisions slightly
contribute to the observed trap loss rate. After correction for
this, a radiative lifetime for CO molecules in the (1,0,1) level of
$2.63\pm0.03$~ms is found, whereas the radiative lifetime for
molecules in the (2,0,2) level is determined as $143\pm4$~ms. The
ratio of these lifetimes is $54.4\pm1.6$, in good agreement with the
ratio of $54.66\pm0.01$ that is expected from the ratio of the $\Omega=1$
character in the respective wavefunctions \cite{Japan}.

To calculate the radiative lifetime of the (1,0,1) level, we use the
model as originally described by James \cite{James:1971}. Essential
in this model are the matrix elements $\langle v_A|h_{A,a}(r)|v_a=0\rangle$ of the $r$-dependent spin-orbit
coupling $h_{A,a}(r)$ and the matrix elements $\langle v_A|\mu_{A,X}(r)|v_X\rangle$ of the $r$-dependent electronic $A^1\Pi - X^1\Sigma^+$ transition dipole moment $\mu_{A,X}(r)$,
together with the energy separation of the spin-orbit coupled
states. We have computed the wave functions $|{v_X}\rangle$,
$|v_A\rangle$, and $|v_a\rangle$, for the vibrational levels of the
$X^1\Sigma^+$, the $A^1\Pi$, and the $a^3\Pi$ state, respectively,
with the sinc-function DVR method \cite{groenenboom:93}. The
diatomic potentials are computed with the RKR program of Ref.
\cite{leroy:04}, taking the required spectroscopic data for
the $X$-state from Ref.\ \cite{lefloch:91} and for the $A$ and $a$
states from Ref.\ \cite{Field:347:1972}.

\begin{figure}[!ht]
\includegraphics[width=0.75\columnwidth]{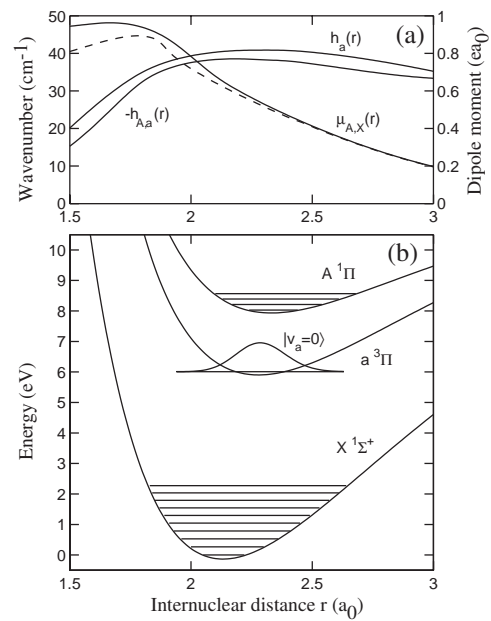}
  \caption{\label{fig:theoretical-results} Panel (a) shows
 the $A-X$ electronic transition dipole moment $\mu_{A,X}(r)$ (in
 $ea_0$) of the present work (the solid line) and from Ref.\ \cite{spielfiedel:99}
  (the dashed line), together with the diagonal [$h_a(r)$] and offdiagonal [$h_{A,a}(r)$]
  spin-orbit coupling (in cm$^{-1}$). Panel (b) shows the RKR-potentials of the
  $X^1\Sigma^+$, the $a^3\Pi$, and the $A^1\Pi$ states of CO
  and the $|v_a=0\rangle$ vibrational wavefunction.}
\end{figure}

James approximated the offdiagonal spin-orbit coupling $h_{A,a}(r)$
as minus the diagonal spin-orbit coupling constant of the $a^3\Pi$
state, $-h_a(r)$ \cite{James:1971}. In the present work we used the
full Breit-Pauli spin-orbit operator and we computed the electronic
wave functions at the internally contracted multireference single
and double excitation (MRCI) level employing molecular orbitals from
a state averaged complete active space multiconfigurational self
consistent field (CASSCF) calculation with the MOLPRO computer
program. The active space consisted of the full valence space
extended with a $\sigma$, a $\pi_x$, a $\pi_y$, and a $\delta_{xy}$
orbital (in all calculations $C_{2v}$ point group symmetry was
used). The $1s$ orbitals were optimized, but not correlated. The
one-electron basis consists of the core-valence correlation
consistent quintuple-zeta (cc-pCV5Z) basis set augmented with tight
$p$ functions with exponents 462.5 and 950.4 for carbon and oxygen,
respectively. The exponents were energy-optimized in atomic
calculations. The accuracy of such an approach was demonstrated by
A. Nicklass \textit{et al.} \cite{nicklass:00}. The transition
dipole moment $\mu_{A,X}(r)$ was also computed at the CASSCF+MRCI
level. The same extended active space was used, but the orbitals for
the $X$ and $A$ states were optimized independently and the cc-pV6Z
basis was used. The results are presented in Fig. 
\ref{fig:theoretical-results}.

Using these results, we computed a radiative lifetime for CO molecules
in the (1,0,1) state of 3.16~ms. In order to understand the $20\%$
discrepancy with the experimental value, we performed an
extensive analysis on the quality of the potentials, and on that of
the calculations of the spin-orbit coupling and transition dipole
moment functions.

James arrived at a calculated value of 2.93~ms for the (1,0,1)
level \cite{James:1971}. We repeated the calculation of James by
employing RKR potentials based on spectroscopic data available in
1971 and found a value of 2.94~ms. With our new RKR potentials and
James' spin-orbit coupling and transition dipole function we obtain
2.93~ms. We therefore do not find the extreme sensitivity to the
potential reported by Sykora and Vidal, and cannot reproduce their
value of 3.41~ms \cite{Sykora:2000}.

We computed the spin-orbit constants of the vibrational levels in
the $a^3\Pi$ state, $\langle {v_a}|h_a(r)|{v_a}\rangle$, because
these can be compared to experimental values. The computed values
for the levels $v_a~=~0-8$ are all between $1.4$ and $1.5$ \% lower
than the spectroscopic values \cite{Field:347:1972}. We checked the
dependence of the spin-orbit coupling on the one-electron basis set
by computing $h_{A,a}$ and $h_a$ at $r=2.3$ $a_0$ with a cc-pCVQZ+p
and a cc-pCVTZ+p basis set. This gives couplings that are about
0.5\%, and 1.8\% higher, respectively, than the value obtained in
the cc-pCV5Z+p basis. We determined the effect of core-valence
correlation on the spin-orbit coupling in a calculation with a
smaller active space and found that it may increase the couplings by
about 0.5\%.

Replacing our $\mu_{A,X}(r)$ with the one computed by Spielfiedel
\textit{et al.} \cite{spielfiedel:99} gives a value of 3.35~ms for
the lifetime of the $(1,0,1)$ level. The \textit{ab initio} method
employed in Ref.\ \cite{spielfiedel:99} is similar to ours. The
difference between their $\mu_{A,X}(r)$ and ours increases at
shorter distances as shown in Fig.\ \ref{fig:theoretical-results}
(a). The main reason for the difference is that we optimize the
orbitals for the $A$ and the $X$ state separately, whereas they
employed orbitals from a state averaged CASSCF calculation.

We compute a total oscillator strength for the $A^1\Pi, v \leq 12 -
X^1\Sigma^+, v=0$ transitions of 0.1764. This is 2.4\% lower than
the value of 0.1807 obtained from high resolution electron energy loss
spectroscopy \cite{Chan:CP170:123}. With $\mu_{A,X}$ from Ref.\
\cite{spielfiedel:99} we find a total oscillator strength of 0.1599, i.e., 11.5\% below the
experimental value.

Another test of $\mu_{A,X}$ is provided by the radiative lifetime
measurements of the $A ^1\Pi$ state by Field \textit{et al.}
\cite{Field:JCP78:2838}. Their deperturbed values for $v=0-7$ are all of 
the same quality and decrease smoothly
from $\tau(v_A=0)=9.9\pm 0.1$~ns to $\tau(v_A=7)=8.95\pm 0.1$~ns. In
our calculations, however, the lifetimes increase from
$\tau(0)=8.4$~ns to $\tau(7)=8.6$~ns. Other \textit{ab initio}
transition dipole functions, e.g., the one of Ref.\
\cite{spielfiedel:99}, also yield lifetimes that increase with
$v_A$. Transition dipole functions fitted to reproduce the measured
$A ^1\Pi$ lifetimes result in computed lifetimes of the $(1,0,1)$
level of 3.5~ms or more.

\begin{table}
\caption{\label{tab:pi-states} The $n ^1\Pi-X ^1\Sigma^+$ transition
dipole moment $\mu_{n,X}$ and the $n ^1\Pi-a ^3\Pi$ spin-orbit
coupling $h_{n,a}$ at $r=2.3$ $a_0$ computed at the CASSCF+MRCI
level (valence CAS extended with 2$\pi_x$ and 2$\pi_y$ orbitals) in an aug-cc-pVTZ basis, the energy
separation $n^1\Pi-a^3\Pi$ ($E_n-E_a$, with Davidson correction) and
the scaled lifetime $\tau_s$ of the (1,0,1) level when the states
$1,..,n$ are included.}
\begin{ruledtabular}
\begin{tabular}{crrr||r}
  $n$ &  $E_n-E_a$ (cm$^{-1}$) & $\mu_{n,X}$ ($ea_0$) & $h_{n,a}$ (cm$^{-1}$) & $\tau_s$ (ms) \\ \hline

1  &    16058 & $   0.542$  & $ -36.3$ &    3.16 \\
2  &    48577 & $  -0.513$  & $   0.1$ &    3.15 \\
3  &    54284 & $   0.369$  & $ -12.9$ &    2.74 \\
4  &    67572 & $   0.016$  & $  17.7$ &    2.76 \\
5  &    70981 & $   0.000$  & $   0.2$ &    2.76 \\
6  &    76952 & $   0.303$  & $   8.9$ &    2.92 \\
7  &    81025 & $  -0.105$  & $  20.8$ &    2.80

\end{tabular}
\end{ruledtabular}
\end{table}

From the results presented so far, we conclude that a model that
takes into account only the $A ^1\Pi$ intermediate state, cannot
explain the observed lifetime of the $(1,0,1)$ level. The main
sources of error are in the $A-a$ spin-orbit coupling and in the
$A-X$ transition dipole moment. It seems unlikely, though, that the
combined effect of these errors on the lifetime is more than 10 \%.
Only $^1\Pi$ states can have both a nonzero transition dipole moment
to the ground state and a nonzero spin-orbit coupling to the
$a^3\Pi_1$ state. Therefore, we made an estimate of the
contributions of higher lying $^1\Pi$ states by calculating the
effective $a-X$ transition dipole moment at $r=2.3$ $a_0$ as a
function of the number of intermediate states, using the result to
scale the value of 3.16~ms obtained above (Table
\ref{tab:pi-states}). The table shows that the higher states could
account for the difference between theory and experiment. We
suspect, however, that taking into account their $r$-dependence and
obtaining convergence with respect to the number of $^1\Pi$ states
will be difficult.

In this paper, we report the electrostatic trapping of metastable CO
molecules. We have exploited the long observation time allowed by
the trap to measure the radiative lifetime of two different
rotational levels in the $a^3\Pi, v=0$ state. These two measurements
are mutually consistent and yield an accurate value of the radiative
lifetime of the $a^3\Pi_1, v=0, J=1$ level of $2.63\pm0.03$~ms. Now
that the radiative lifetime of this level is known with
unprecedented precision, discrepancies with earlier calculated
values have become apparent. This prompted us to perform more
detailed calculations for this level. When only spin-orbit coupling
of the $a^3\Pi$ state with the $A^1\Pi$ state is included, we
compute a lifetime of 3.16~ms. We show that obtaining agreement
between theory and experiment will require the calculation of the
contribution of higher $^1\Pi$ states.

We acknowledge helpful discussions with A. van der Avoird and R. W.
Field. M. M. is grateful to the Academy of Finland for financial
support.


\end{document}